\pdfoutput=1
\documentclass[preprint,12pt]{elsarticle}

\usepackage{mathtools}
\usepackage{amssymb}
\usepackage{amsmath}
\usepackage[T1]{fontenc}
\usepackage{float}
\usepackage{xcolor}

\usepackage{siunitx}
\usepackage{tikz}
\usepackage{upgreek}

\usepackage{graphicx} 
\usepackage{booktabs}
\usepackage{tabularx}

\usepackage{lineno}

\usepackage{subcaption}
\usepackage{mwe}
\usepackage{pdfpages}

\usepackage{hyperref}
\hypersetup{colorlinks,allcolors=black}
\usepackage{comment}
\usepackage{physics}
\usepackage{accents}

\graphicspath{{figures/}{figures/dEdx/}}

\journal{Nuclear Instruments and Methods}

\begin{document}

\begin{frontmatter}





\title{Characterization of the Electronic Noise in the Readout of Resistive Micromegas in the High-Angle Time Projection Chambers of the T2K Experiment}

\author[saclay]{D.~Attié}
\author[lpnhe]{P.~Billoir}
\author[padova]{G.~Bortolato}
\author[saclay]{S.~Bolognesi}
\author[Bari]{N.~F.~Calabria}
\author[saclay]{D.~Calvet}
\author[Bari]{M.~G.~Catanesi}
\author[padova]{G.~Collazuol}
\author[saclay]{P.~Colas}
\author[padova]{D.~D'Ago}
\author[saclay]{T.~Daret}
\author[saclay]{A.~Delbart}
\author[lpnhe]{J.~Dumarchez}
\author[saclay]{S.~Emery-Schrenk}
\author[padova]{M.~Feltre}
\author[padova]{C.~Forza}
\author[lpnhe]{A.~N.~Gacino Olmedo}
\author[lpnhe]{C.~Giganti}
\author[lpnhe]{M.~Guigue}
\author[saclay]{G.~Eurin}
\author[saclay]{S.~Hassani\fnref{fnref1}}
\author[saclay]{D.~Henaff}
\author[saclay]{S.~Joshi}
\author[saclay]{J.~F.~Laporte\fnref{fnref2}}
\author[padova]{S.~Levorato}
\author[IFAE]{T.~Lux}
\author[Bari]{L.~Magaletti}
\author[padova]{L.~Mareso}
\author[padova]{M.~Mattiazzi}
\author[IFAE]{E.~Miller}
\author[lpnhe]{B.~Popov}
\author[Bari]{C.~Pastore}
\author[IFAE]{C.~P\'io}
\author[Bari]{E.~Radicioni}
\author[lpnhe]{L.~Russo}
\author[aachen]{S.~Roth}
\author[lpnhe]{W.~Saenz Arevalo}
\author[padova]{L.~Scomparin}
\author[lpnhe]{Ph.~Schune}
\author[aachen]{D.~Smyczek}
\author[aachen]{J.~Steinmann}
\author[aachen]{N.~Thamm}
\author[lpnhe]{U.~Virginet}
\author[saclay]{G.~Vasseur}
\author[IFAE]{M.~Varghese}
\author[Bari]{V.~Valentino}
\author[lpnhe]{M.~Zito}
\address[saclay]{IRFU, CEA, Universit\'e Paris-Saclay, Gif-sur-Yvette, France}
\address[padova]{INFN Sezione di Padova and Universit\`a di Padova, Dipartimento di Fisica e Astronomia, Padova, Italy}
\address[Bari]{INFN , Universita’ e Politecnico di Bari}
\address[lpnhe]{LPNHE Paris, Sorbonne Universit\'e, CNRS/IN2P3, Paris 75252, France}
\address[aachen]{III. Physikalisches Institut, RWTH Aachen University, Aachen, Germany}
\address[IFAE]{Institut de Física d’Altes Energies (IFAE) - The Barcelona Institute of Science and Technology (BIST), Campus UAB, 08193 Bellaterra (Barcelona), Spain}

\cortext[cor1]{Corresponding authors}
\fntext[fnref1]{samira.hassani@cea.fr}
\fntext[fnref2]{jean-francois.laporte@cea.fr}

\begin{abstract}
The two high-angle Time Projection Chambers of the T2K experiment are equipped with a new readout system based on resistive Micromegas detector technology, and utilize custom-made electronics based on AFTER chips for signal processing. This study analyzes and characterizes the electronic noise of the detector readout chain to develop a comprehensive noise model. The model enables the generation of Monte Carlo simulations to investigate systematic effects in signal processing. The analysis is based on data collected from 32 resistive Micromegas detectors, recorded without zero suppression. All detectors exhibit a quasi-identical and time-stable noise level. The developed analytical model accurately describes the observed noise, and derived Monte Carlo simulations show excellent agreement with experimental data.
\end{abstract}


\end{frontmatter}
\newpage
\tableofcontents
\newpage
\section{Introduction}
The T2K (Tokai to Kamioka)~\cite{T2K:2011qtm} is a long baseline neutrino oscillation experiment in Japan, which conducts measurements of neutrino oscillation parameters by generating a highly intense muon (anti) neutrino beam peaked at 600 MeV at the \mbox{J-PARC} facility. This beam is measured 280~m from its point of origin by a set of near detectors (ND280), positioned prior to oscillations, with the aim of monitoring and constraining systematic uncertainties associated with the neutrino flux and interaction models. Subsequently, the far detector, Super-Kamiokande, which is located 295~km away, is responsible for detecting the disappearance of muon (anti)neutrinos and the appearance of electron (anti)neutrinos within the beam.

\begin{figure}[ht]
    \begin{center}
        \includegraphics[width=0.4\textwidth]{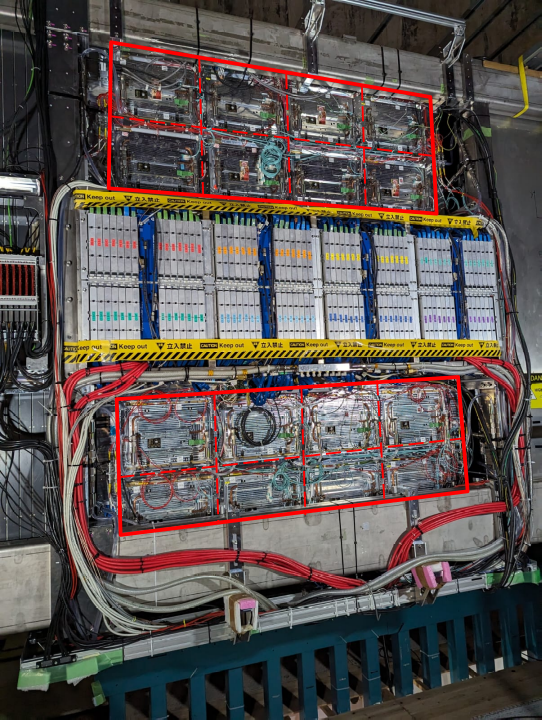}
        \caption{ The new sub-detectors installed in ND280 include the finely segmented active target (Super-FGD), positioned between the top and bottom high-angle time projection chambers. The red lines indicate the ERAM detector boundaries.
        \label{fig:ERAM_production}}
    \end{center}
\end{figure}

The ND280 has recently undergone an upgrade, which involves the integration of a suite of subdetectors positioned at the upstream section of the existing ND280 setup as shown in Fig.~\ref{fig:ERAM_production}. This upgraded configuration features a finely segmented active target known as the Super-FGD, positioned between two high angle time projection chambers (HA-TPCs), all enclosed by a time-of-flight detector (ToF). The installation of these new detectors at J-PARC was completed in May 2024, and the neutrino data were collected during June 2024.

\begin{figure}[ht]
    \begin{center}
        \includegraphics[width=0.45\textwidth]{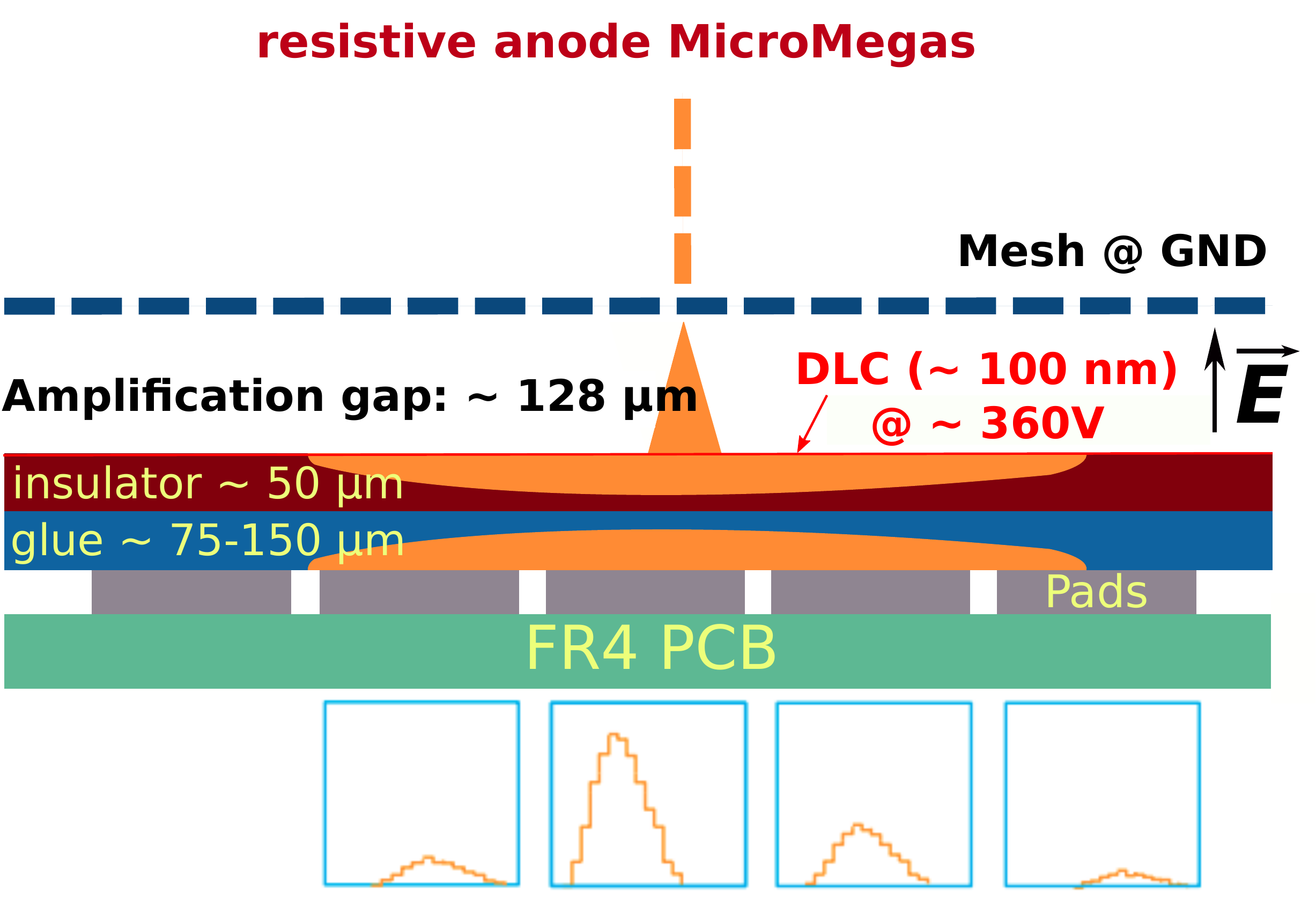}
        \caption{Operating principle of Resistive Micromegas. The various elements of the detector are highlighted. The glue thickness is adapted to keep a constant RC for a given detector~\cite{Ambrosi:2023smx}. 
        \label{fig:ERAM_operating_principle}}
    \end{center}
\end{figure}

The field cage of the HA-TPC features a new design that minimizes dead space and maximizes the tracking volume. It is equipped with a new readout based on the innovative resistive Micromegas technology~\cite{Attie:2022smn}. Fig.~\ref{fig:ERAM_operating_principle} illustrates the concept of resistive Micromegas. In the resistive anode design, charge dispersion is achieved by applying a resistive foil of diamond-like carbon (DLC) over the pad plane. As the initial charge spreads across multiple pads, it allows for improved position resolution without the need to reduce the size of the readout pads, making this a compact and cost-effective technology. The resistive foil on the anode slows down the charge dissipation process, thereby reducing the frequency and intensity of sparks without the need for additional spark protection measures. A novel high-voltage powering scheme is utilized in resistive anode Micromegas operation, where the mesh is grounded, and the anode is set to a positive voltage. This approach enhances the safety of detector handling and improves the uniformity of the electric field. Due to this high-voltage scheme, the current Micromegas design is referred to as Encapsulated Resistive Anode Micromegas, or "ERAM" for short.

The ERAM module consists of a resistive Micromegas detector attached to an aluminum frame and the associated readout electronic cards, inserted on its backside. The $42 \times 34  \text{cm}^2$ detector features 1152 pads, each measuring $11.18 \times 10.09  ~\text{mm}^2$, arranged in a matrix with 36 pads in the $x$ direction and 32 pads in the $y$ direction. The pad plane is covered by a resistive layer made of insulated $50~\mu$m Apical polyimide foil, pressed with $150~\mu$m glue, onto which DLC is deposited. To ensure charge dispersion over at least two pads, a DLC surface resistivity $R$ of approximately $400 ~\text{k}\Omega/\square$ was selected. A schematic cross-section of the ERAM detector and its specifications are shown in Fig.~\ref{fig:ERAM_operating_principle}. This charge spreading is driven by the $RC$ constant of the resistive stack. In this configuration, $R$ is the surface resistivity of the DLC foil and $C$ the capacitance per unit area of the stack governed by the glue thickness~\cite{Ambrosi:2023smx}.
The new readout electronics is based on the AFTER chip~\cite{Baron:2008zza}. The detector characterization and performance are detailed in Refs.~\cite{Attie:2022smn, Ambrosi:2023smx, Attie:2021yeh, Attie:2019hua}.\\

The goal of this paper is to analyze and characterize the noise in the ERAM detectors and front-end electronics to develop a comprehensive noise model. This model will facilitate the generation of Monte Carlo simulations for investigating the systematic effects in signal processing. The analysis is based on data collected from 32 ERAMs, evenly distributed between the top and bottom HA-TPCs, recorded without zero suppression.\\

The paper is organized as follows: Section~\ref{sec:Electronic Response} provides a brief description of the AFTER chip and its electronic response. Section~\ref{sec:Noise Characteristics} presents a detailed analysis of noise characterization and Fast Fourier Transform analysis. The various components of the developed analytical model are then introduced, followed by a fit to the data in Section~\ref{sec:FFT_analysis_section}. Section~\ref{sec:MCproductions} discusses Monte Carlo simulations generated using the analytical model and compares their results with the data. Finally, the conclusions are presented in Section~\ref{sec:conclusion}.

\label{sec:introduction}
\section{Electronic Response}
\label{sec:Electronic Response}
The new readout electronics is based on the AFTER chip~\cite{Baron:2008zza}. The sampling frequency is set to $f_e =25 \;{\rm MHz}$ and a peaking time of either 200 or 412~ns is chosen. The selected input charge dynamic range is 120 fC and digitization is done with a resolution of 12 bits. Each AFTER chip reads 72 detector channels covering a $9 \times 8$ array of pads. A Front End Card (FEC) houses eight AFTER chips and performs the digitization of 576 pad signals. Two FECs are required to read a complete ERAM module. These cards are directly plugged at the back of the PCB of the detector. A Front End Mezzanine card (FEM) mounted above the two FECs synchronizes signal sampling, collects digitized data, and ensures transfers to the back-end electronics. An aluminum carapace covers each FEC and FEM to provide shielding against electromagnetic interference and to conduct the generated heat to copper pipes that circulate cold water. Finally, data from all 16 ERAM modules of a HA-TPC are transmitted via optical fibers to a custom-made board known as the Trigger and Data Concentrator Module (TDCM)~\cite{Calvet:2018lac}. \\

\begin{figure}[hbt!]
  \centering
  \includegraphics[width=0.95\textwidth]{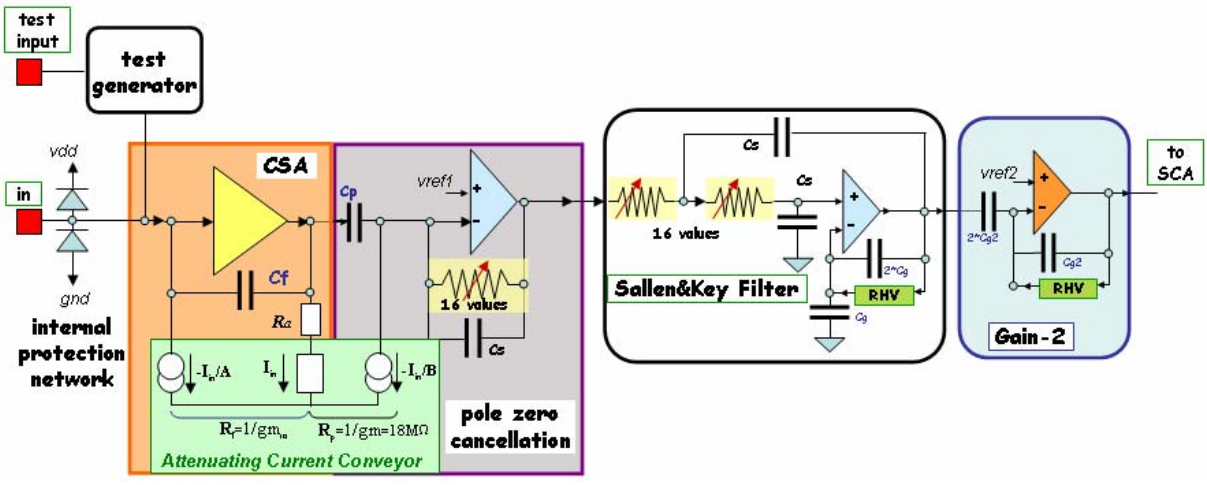}
  \caption{Architecture of one channel of the AFTER chip~\cite{Baron:2008zza}.}
  \label{fig:AFTERFIG}
\end{figure}

The architecture of the AFTER chip is shown in Fig.~\ref{fig:AFTERFIG}. It is composed  of a charge integration stage (CSA), a pole zero compensation stage, a Sallen \& Key filter, an amplifier, and an analog memory composed of a 511-cell switched capacitor array (SCA). Upon an external trigger signal, the content of the SCA is frozen and all cells are sequentially digitized by an external Analog-Digital converter (ADC).

The input current being $I_{in}(t)$, the linearity of the electronics imposes that the
output ADC signal, 
$ADC(t)$ is the convolution product 
\begin{equation}
  ADC(t) = I_{in} \circledast ADC^{D}
    \label{eq:convformula}
\end{equation}
where the  function $ADC^{D}(t)$ is the AFTER chip response to a Dirac pulse current, $I_{in}(t)=\delta(t)$.

The transfer function, which describes the output amplitude as a function of the input signal's frequency, corresponds to the Laplace transform of the pulse response. The transfer function of the AFTER chip is obtained by multiplying the transfer function of its individual stages, assuming ideal behavior for each.
By assuming perfect pole zero compensation, 
we derive the Laplace transfer function of the AFTER chip, as given by
\begin{equation}
H(s)=\frac{V_{\text {out }}(s)}{I_{\text {in }}(s)} \propto  \frac{1}{1+\left(\frac{s}{\omega_{s}}\right)} \frac{1}{\left(\frac{s}{\omega_{s}}\right)^{2}+\frac{1}{Q}\left(\frac{s}{\omega_{s}}\right)+1}  
\label{Eq:3.2}
\end{equation}
where $I_{\text {in }}$ is the input current and $V_{\text {out }}$ is the output voltage.\\
The parameters $\omega_{s}$ and $Q$ are the natural frequency of the electronics  and the quality factor. 
Their values are determined by the circuit's resistance and capacitance values.
Both parameters have been measured through a fit to electronics calibration data~\cite{Ambrosi:2023smx} yielding the values $Q=0.6368$ and $\omega_{s}= \frac{2}{409.88 \;{\rm ns}}$.

\begin{figure}[hbt!]
     \centering
     \begin{subfigure}[b]{0.49\textwidth}
         \includegraphics[width=1.\textwidth]{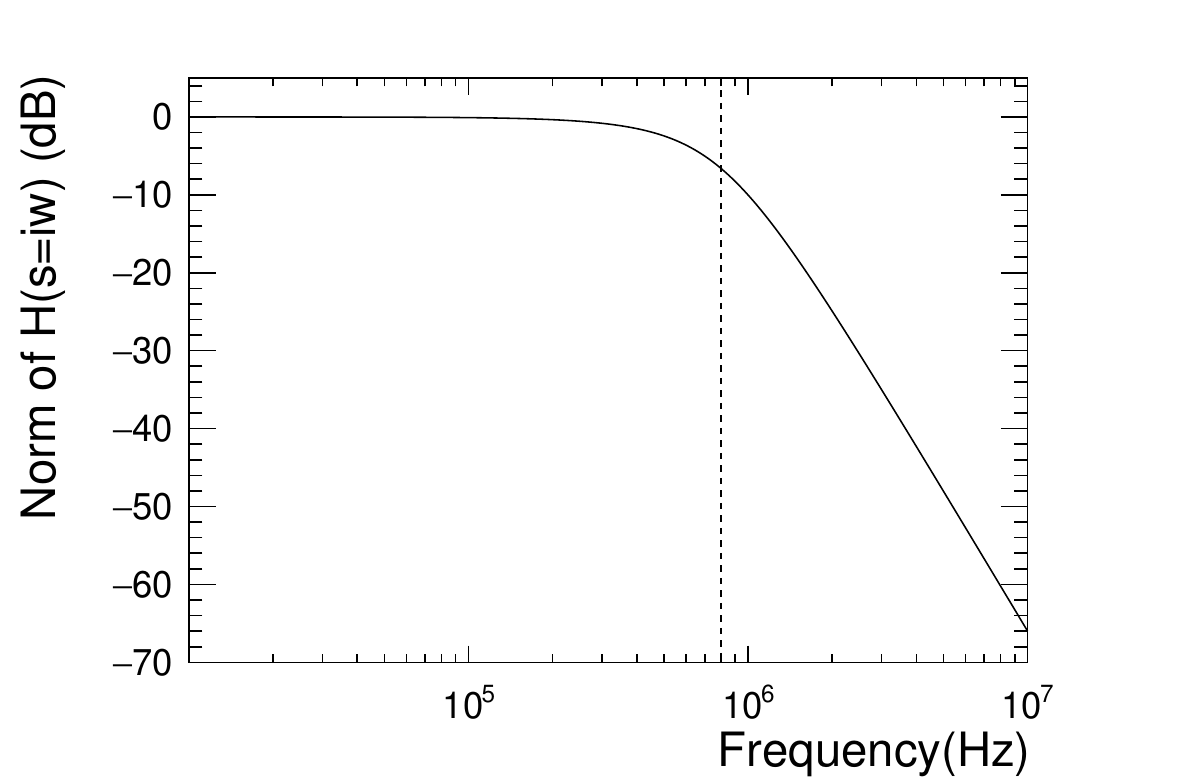}
         \caption{Norm of the frequency function $H(s=iw)$.}
         \label{fig:p16}
     \end{subfigure}
     \hfill
     \begin{subfigure}[b]{0.49\textwidth}
         \includegraphics[width=1.\textwidth]{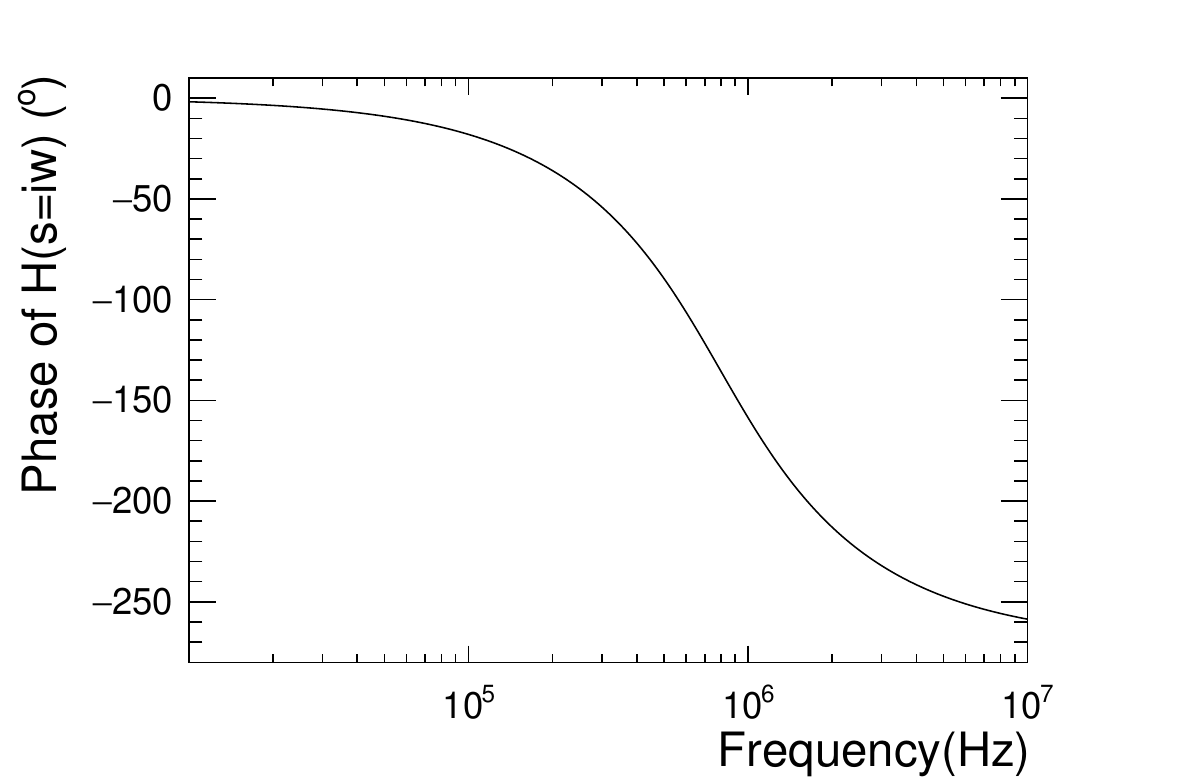}
         \caption{Phase of the frequency function $H(s=iw)$.}
         \label{fig:p16p17}
     \end{subfigure}
        \caption{Norm and Phase of the frequency function $H(s=iw)$ for $Q=2/3$ and $\omega_{s} = \frac{1}{200\;{\rm ns}}$.}
        \label{fig:fig_3.2}
\end{figure}

The norm and the phase of the frequency function $H(s=iw)$ are shown
in Figs.~\ref{fig:p16p17} for the typical value of $Q=2/3$
and for $\omega_{s} = \frac{1}{200\;{\rm ns}}$ which gives a peaking time of the Pulse Response at $T_p \sim\frac{2}{ \omega_{s}}=400 \;{\rm ns}$.
As it can be seen in Fig.~\ref{fig:p16}, the Bode amplitude plot displays a low-pass filter behavior which strongly attenuates the frequencies above $f_s= \frac{\omega_{s}}{2\pi} \sim 0.8\;{\rm MHz}$.

\begin{figure}[hbt!]
  \centering
  \includegraphics[width=0.5\textwidth]{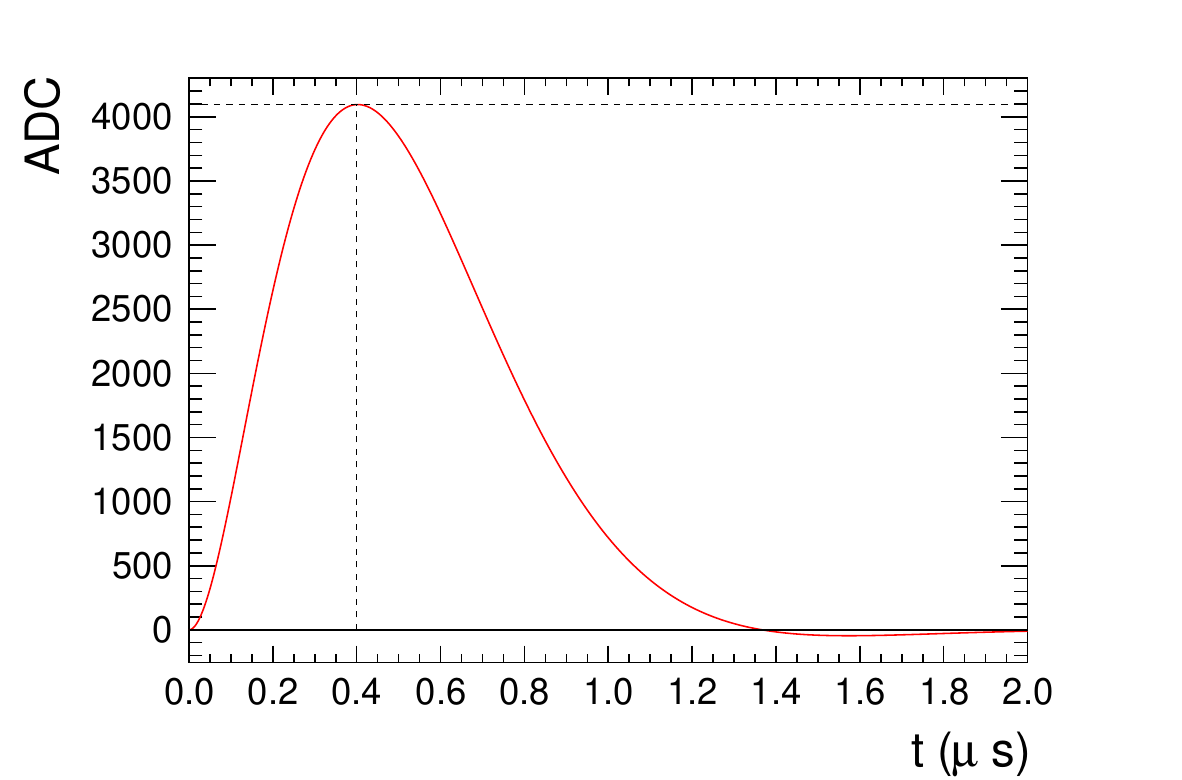}
  \caption{Response to a current pulse for $Q=2/3$ and $\omega_{s}=\frac{1}{200\;{\rm ns}}$.
  }
  \label{fig:p18}
\end{figure}

From Eq.~\eqref{Eq:3.2} it comes that the response to a Dirac current pulse, prior to the late stage of discretization, is of the form
\begingroup
\small
\begin{equation}
ADC^{D}(t) 
\propto
e^{ -\omega_{s}t }
+
   e^{ -   \frac{\omega_{s}t}{2 Q}}    
  \left[
  \sqrt{ \frac{ 2Q- 1 }{ 2Q +1 } }
  \sin \left( \frac{\omega_{s}t }{2} \sqrt{4-\frac{1}{Q^2}} \right)
-
  \cos \left( \frac{\omega_{s}t }{2} \sqrt{4-\frac{1}{Q^2}} \right)
  \right]
\label{equ:elecResponse}
\end{equation}
\endgroup
As the $Q$ factor increases, the filter exhibits increased "ringing" at a single frequency. In fact, taking the limit $Q \to \infty$ in Eq.~\eqref{equ:elecResponse} reveals that, aside from a transient decaying exponential, the signal consists solely of sinusoidal components at frequency 
$\omega_{s}$.
However for finite $Q$, the oscillations are damped and for the value $Q=2/3$ typical of the AFTER chip, 
the pulse response reduces mainly to the  first oscillation as illustrated by Fig.~\ref{fig:p18}.\\

Despite the significant simplifications applied to the various stages of the actual AFTER chip in the computations presented here, the function in Eq.~\eqref{equ:elecResponse} has proven to be sufficiently accurate and it has been extensively used in the analysis of ERAM detector data~\cite{Ambrosi:2023smx}.

\section{Noise Characteristics}
\label{sec:Noise Characteristics}
\begin{figure}[hbt!]
     \centering   
     \begin{subfigure}[b]{0.49\textwidth}
         \includegraphics[width=1.\textwidth]{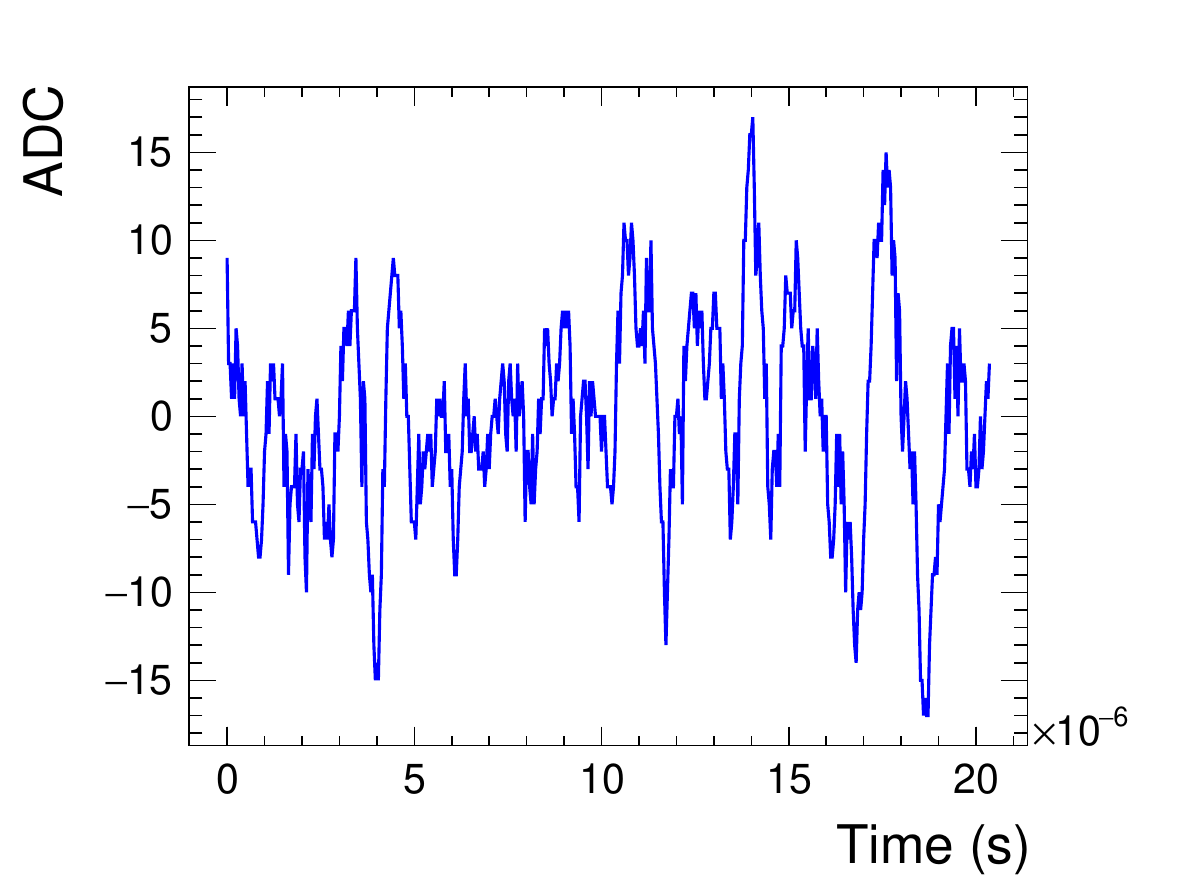}
          \caption{}
         \label{fig:NoiseWF}
     \end{subfigure}
     \hfill
      \begin{subfigure}[b]{0.49\textwidth}
         \includegraphics[width=1.\textwidth]{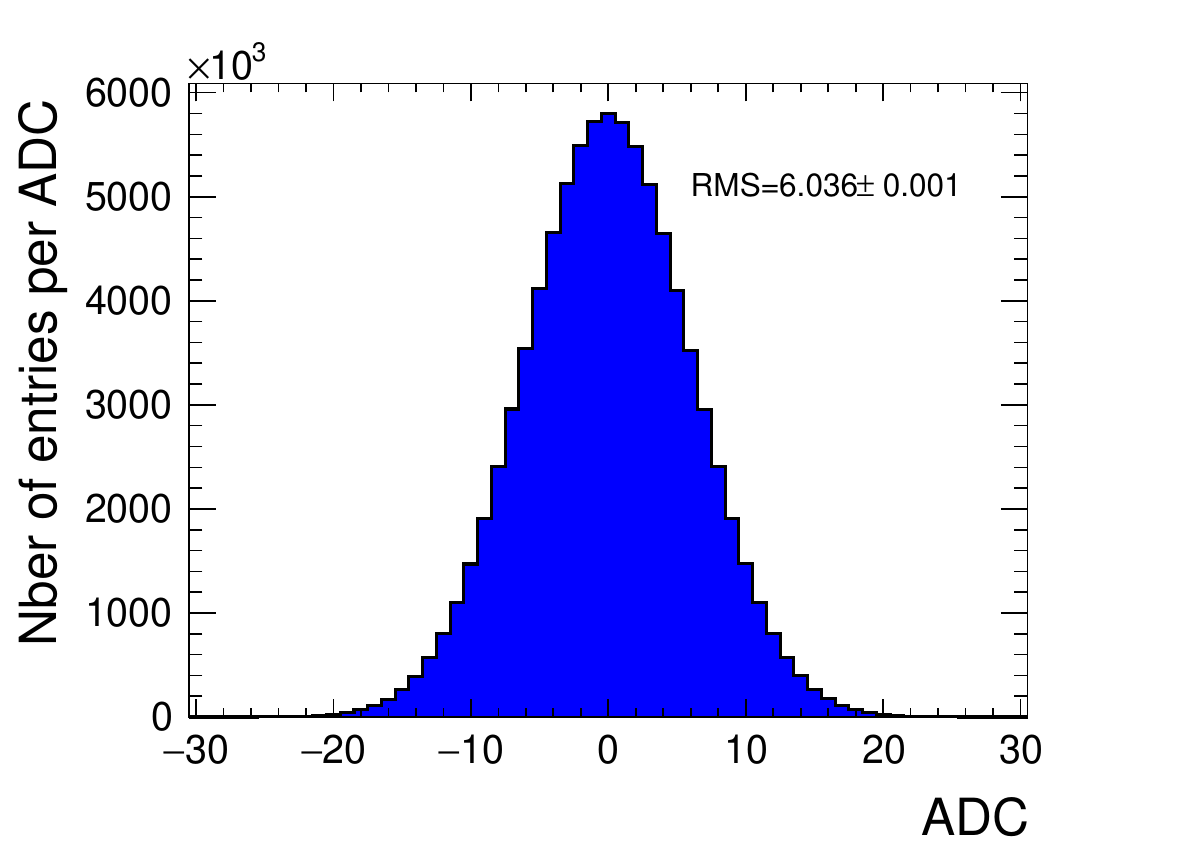}
         \caption{}
         \label{fig:ADC(15)}
     \end{subfigure}    
        \caption{Example of one record of the noise waveform 
        for a pad using a peaking time
         of $T_p =412~\mathrm{ns}$, and a sampling frequency $f_s = 25~\mathrm{MHz}$        (left). RMS distribution for all ERAMs except ERAM 29 (right).
        }
        \label{fig:NoiseFFT}
\end{figure}

The objective of this study is to analyze and understand the noise characteristics of the HA-TPC detectors, with the aim of developing a noise model. The analysis is based on data collected from 32 ERAMs, distributed across the top and bottom HA-TPCs,  recorded without zero suppression using a peaking time of $T_p =412~\mathrm{ns}$, and a sampling frequency $f_s = 25~\mathrm{MHz}$. The signal is recorded and digitized every $40~\mathrm{ns}$, and each waveform spans 511 time bins, covering a time window of $20.4~\mathrm{\mu s}$. Each dataset consists of 10 waveforms per pad. 
Fig.~\ref{fig:NoiseWF} shows a single recording of the noise waveform from a pad. The noise RMS value is 6 ADC units as shown in Fig.~\ref{fig:ADC(15)}. 
This amounts to an
Equivalent Noise Charge (ENC)
of $\sim 1100$ electrons~\footnote{The noise level of an AFTER channel with no external connection is 
$\sim 1.5 $ ADC unit, i.e $\sim 275$ electrons
while the noise level of a FEC with no detector is 
$\sim 3.2$ ADC unit, i.e $\sim 585$ electrons.
}.
The largest amplitudes are observed at low frequencies, typically around $1~\mathrm{MHz}$, suggesting effective suppression of higher frequencies. However, a small high-frequency noise component is also present.  For a more quantitative characterization, the frequency spectrum of the signal can be analyzed using a Fast Fourier Transform.

\section{Fast Fourier Transform Analysis}
\label{sec:FFT_analysis_section}
\begin{figure}[hbt!]
\centering 
\includegraphics[scale=0.5]{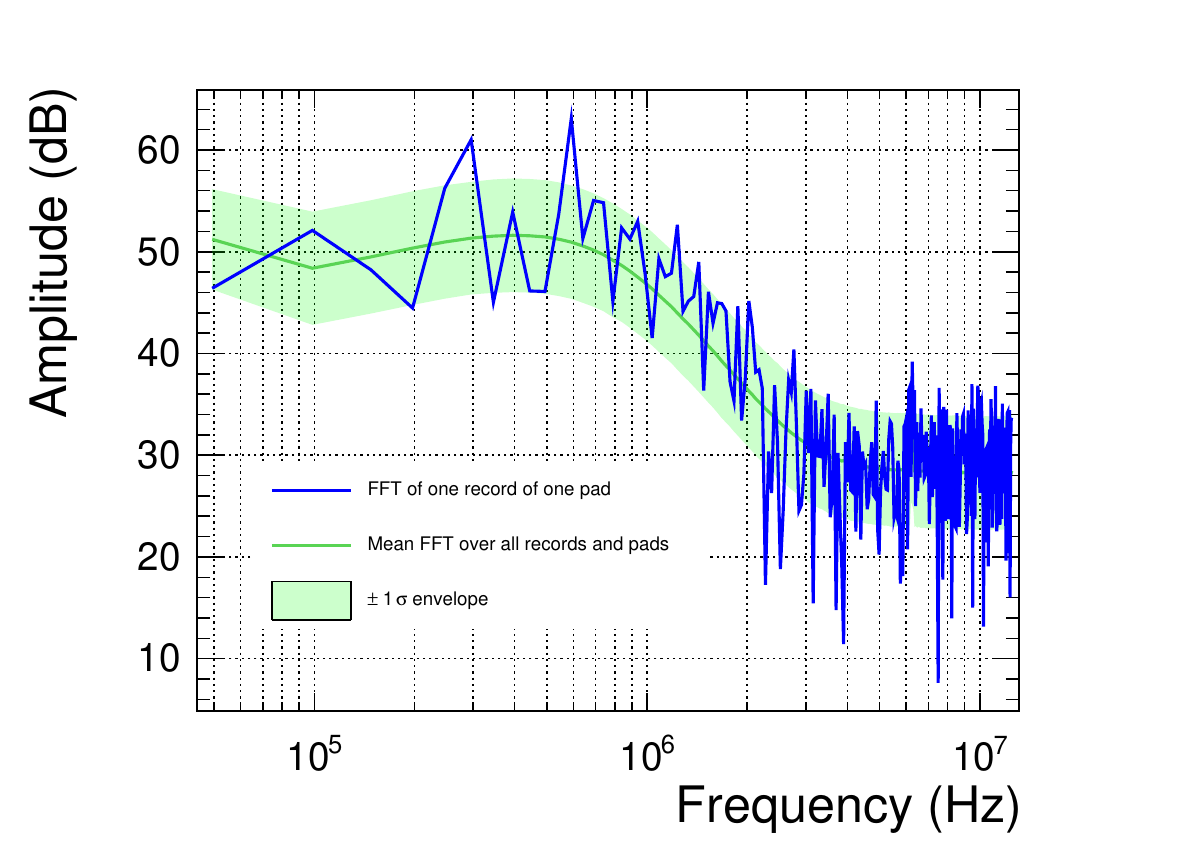}
\caption{Example of the FFT of one record and mean over 10 records for each pad of each ERAM, except ERAM $29$.
Blue curve : FFT of one pad for one record;
Green curve: Mean FFT over all pads and all records;
Green area : $\pm$ one RMS for all the FFT of all pads and all records.
}
\label{fig:FFTexample}
\end{figure}

 Fig.~\ref{fig:FFTexample} presents an example Fast Fourier Transform (FFT) of a pad waveform, where the chaotic spectrum makes it difficult to extract meaningful features. To address this, our analysis focuses on averaged data, specifically the mean FFT with an envelope representing $\pm$ one RMS fluctuations around the mean shown in Fig.~\ref{fig:FFTexample}.
 
\begin{figure}[hbt!]
     \centering
     \begin{subfigure}[b]{0.49\textwidth}
         \includegraphics[width=1.\textwidth]{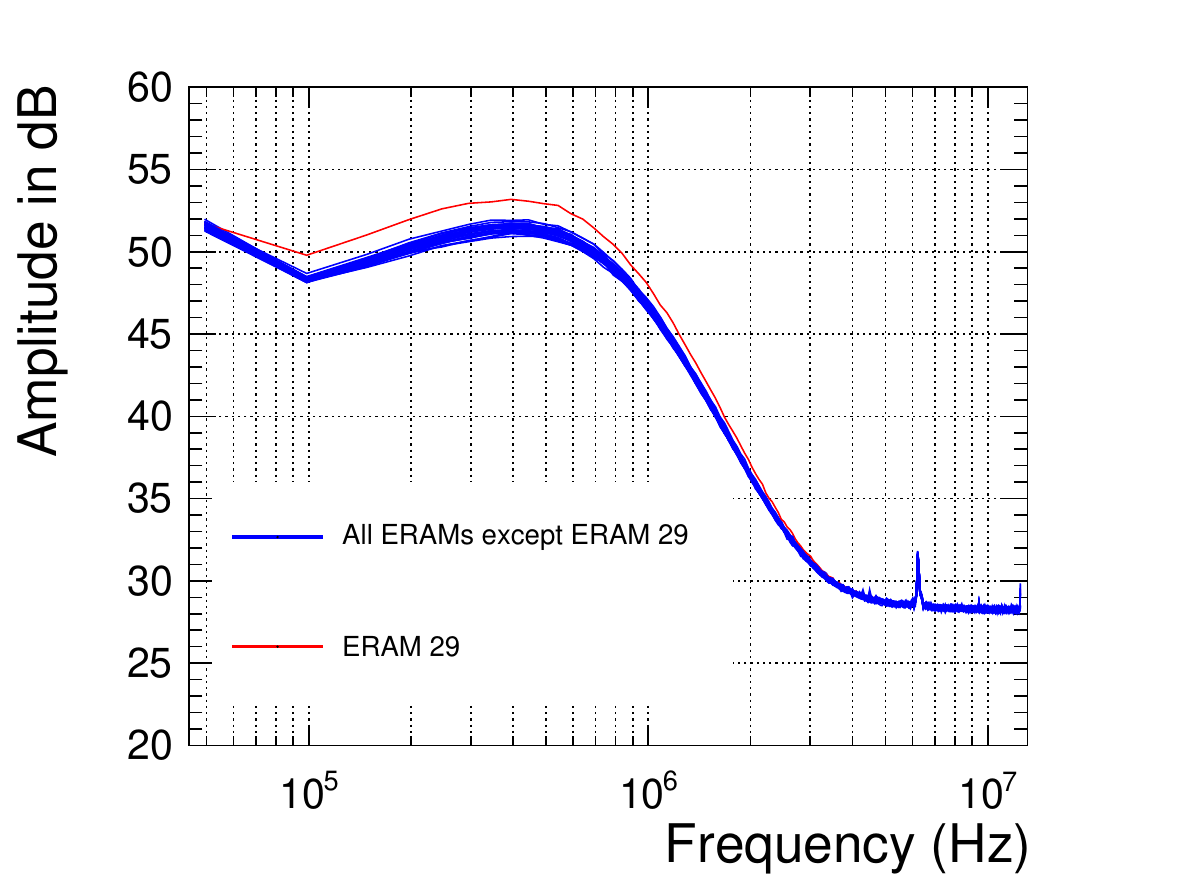}
         \caption{}
         \label{fig:ERAMNoiseTopBottomHATPC}
     \end{subfigure}
     \hfill
     \begin{subfigure}[b]{0.49\textwidth}
         \includegraphics[width=1.\textwidth]{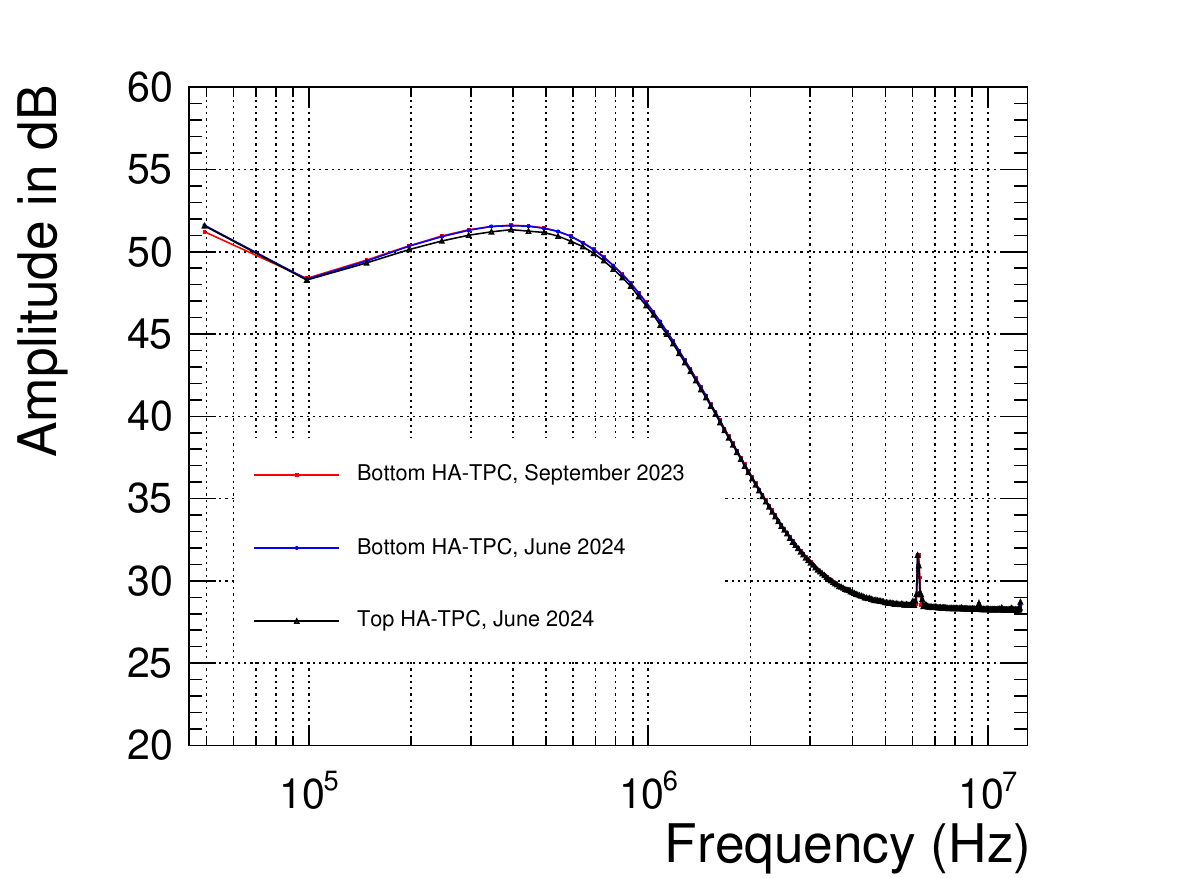}
         \caption{}
         \label{fig:NoiseTimeEvolution}
     \end{subfigure}
        \caption{
        Mean FFTs over all records of all pads of each of the 32 ERAMs including both bottom and top TPC (left) 
        and evolution in time over nine months 
        of the mean FFT over all pads and all ERAMS except ERAM 29 for bottom TPC (right).
        }
        \label{fig:Gain_02_QA_OffCenter}
\end{figure}

Fig.~\ref{fig:ERAMNoiseTopBottomHATPC} shows the mean FFT across all the 32 ERAMs in the top and bottom HA-TPC. The spectra of the different ERAMs are remarkably similar except the spectra of ERAM 29, which exhibits a higher noise amplitude. 
All ERAMs were produced using DLC foils with a surface resistivity of approximately $400~\text{k}\Omega/\square$ and a glue thickness of $150~\mu\text{m}$, except for ERAM~29, which features half the glue thickness and half the surface resistivity. The lower resistivity of its DLC foil is compensated by reducing the Kapton and glue layer thicknesses, in order to maintain a stable and consistent $RC$ value across the 32 produced ERAMs. The higher capacitance of ERAM~29 leads to greater noise : the ADC distribution analysis shows an RMS of 7 ADC units for ERAM~29, compared to the typical 6 ADC units observed in other ERAMs.

Despite minor variations due to load capacitance, the precise mechanical construction of the ERAM modules ensures consistent noise levels across 31 of them, minimizing the need for individual characterization. The RMS of the variation of the FFT across all 32 ERAMs is within 5\% of the mean value. Additionally, a comparison between an earlier dataset from September 2023 and a more recent one from June 2024, collected with the bottom HA-TPC, confirms that the noise remains stable over nine months, as illustrated in Fig.~\ref{fig:NoiseTimeEvolution}. Overall, all ERAMs exhibit a quasi-identical and time-stable noise level.

\subsection{Anatomy of the FFT Spectrum}
A noise study is conducted by constructing an analytical function to fit the data and extract relevant features.
Examining the mean FFT across all ERAMs reveals that most of the noise is concentrated below 1~$\mathrm{MHz}$. As shown in Fig.~\ref{fig:Gain_02_QA_OffCenter}, the mean noise spectrum can be divided into four main regions :
\begin{itemize}
\item Central region ($0.1$  $\mathrm{MHz}$ to $3$ $\mathrm{MHz}$): this part closely resembles the Bode amplitude plot characteristic of the AFTER chip signal (see Fig.~\ref{fig:p16}).
\item Low-frequency region (below $0.1$ $ \mathrm{MHz}$): a decreasing slope is observed in this range.
\item High-frequency plateau (above $4$ $ \mathrm{MHz}$): the spectrum levels off at approximately $\sim 28.5 \mathrm{~dB}$.
\item Distinct peak ($\sim 6$ $\mathrm{MHz}$): 
varying the sampling frequency  $f_e$ has revealed that this peak appears systematically at $f_{p}=\frac{f_{e}}{4}$.
\end{itemize}
The shape of the spectrum in its central region (from $0.1 \mathrm{MHz}$ to $3 \mathrm{MHz}$) is understood as the effect of the AFTER chip convoluted with some random current. It is therefore natural to describe this central part with the function
\begin{equation}
     I(f) \times |H(s=i2\pi f)| 
     \nonumber
\end{equation}    
where $|H(s=i2\pi f)|$  is the norm of the transfer function of the AFTER chip (cf Eq.~\eqref{Eq:3.2})
and 
$I(f)$ is the frequency spectrum of the input noise current.
To  fit the lowest-frequency region of the spectrum, an additional $\frac{1}{f^2}$  component is introduced, while a constant term is added to account for the highest-frequency region.

\begin{figure}[hbt!]
\centering
\includegraphics[scale=0.5]{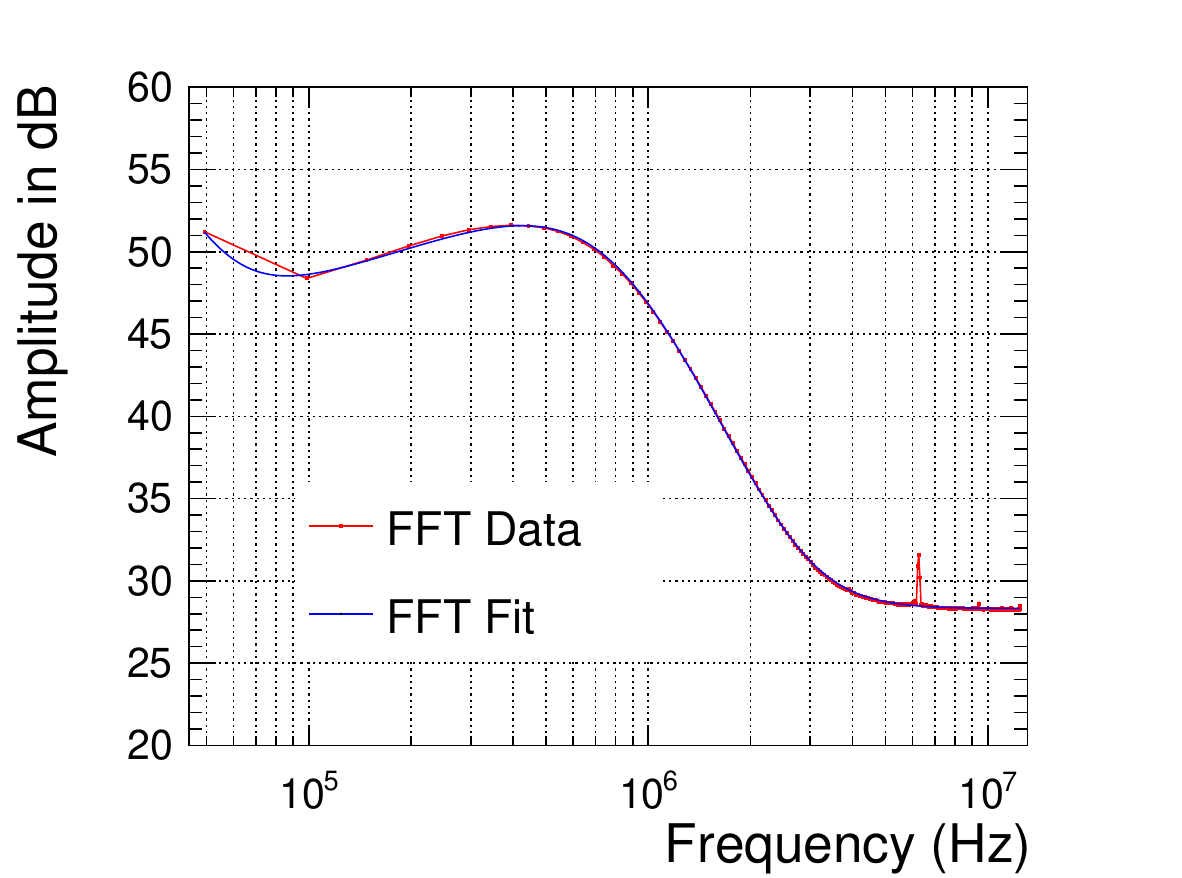}
\caption{Results of the analytical fit model to mean FFT in data.}
\label{fig:fig.5.2}
\centering
\end{figure}

Ignoring the peak at $\frac{f_e}{4}$, the mean of the FFT amplitudes is therefore
fitted with a three components function:
\begin{align}
 ADC\left(f \right)=  
 \sqrt{
 C_{1}^{2}\left(f  \right)
 +
 C_{2}^{2}\left(f  \right)
 +
 C_{3}^{2}\left(f  \right)} 
\nonumber
\end{align}
where  
\begin{gather}
C_{1}\left(f  \right)  =\frac{A_{1}}{\left(f / f_{f}\right)^{2}} 
\text{ , }
C_{2}\left(f  \right)   =I\left(f  \right)\left|H\left(s=i 2 \pi f  \right)\right| 
\text{ and }
C_{3}\left(f  \right)  = A_{3} 
    \nonumber
\end{gather}
with  
\begin{equation}
I\left(f  \right)=A_{2} \sqrt{\alpha_{2}^{2}\left(f / f_{a}\right)^{2}+\left(f / f_{a}\right)+\gamma_{2}^{2}} 
\nonumber
\end{equation}
The frequencies $f_f$ and $f_a$, introduced for convenience, are   $f_f = f_e/511$ 
and $f_a=0.1 \;{\rm MHz}$.
The parameters of the fit are $A_1$, $A_2$, $\alpha_2$, $\gamma_2$ and $A_3$.
Using  the values  of $Q$ and $w_s$ obtained from the electronic calibration~\cite{Ambrosi:2023smx}, the parameters extracted from the fit of the noise data are listed in Table \ref{table:1}. Fig.~\ref{fig:fig.5.2} shows a good agreement between the average FFT and the analytical fit.

\begin{table}[h]
\centering
\begin{tabular}{|c|c|c|c|c|}
\hline \multicolumn{5}{|c|}{ Parameters } \\
\hline $A_{1}(d B)$ & $A_{2}(d B)$ & $\alpha_{2}$ & $\gamma_{2}$ & $A_{3}(d B)$ \\
\hline $49 \pm 4$ & $46 \pm 1$ & $0.16 \pm 0.04$ & $0.8 \pm 0.3$ & $28.3 \pm 0.2$ \\
\hline
\end{tabular}
\caption{The parameter values are obtained from fitting the mean of the FFT amplitudes in the data, ignoring the peak at $\frac{f_e}{4}$.}
\label{table:1}
\end{table}

\subsection{Deterministic Contribution}  
\begin{figure}[hbt!]
\centering
\includegraphics[scale=0.5]{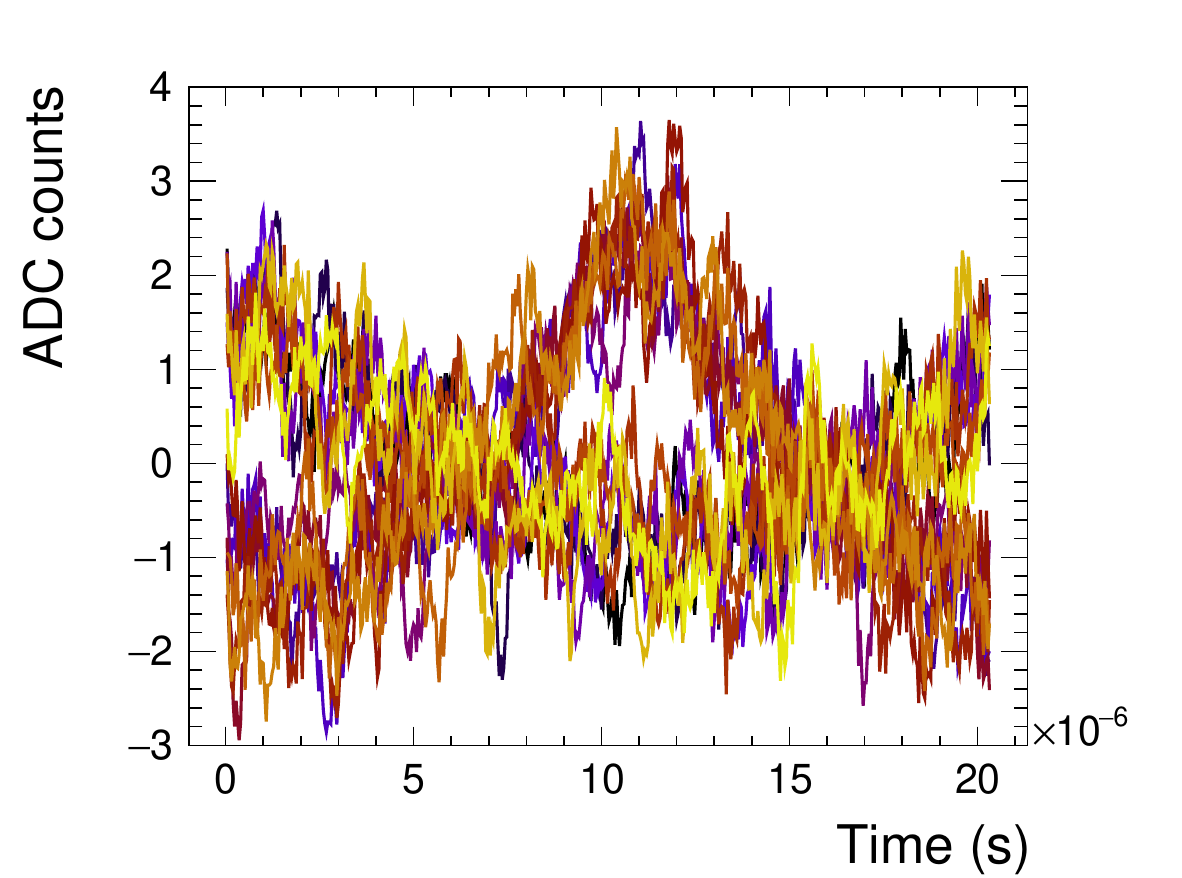}
\caption{Mean waveforms of few pads in the ERAMs, averaged over all records. }
\label{fig:5.3}
\centering
\end{figure}

\begin{figure}[hbt!]
     \centering
     \begin{subfigure}[b]{0.49\textwidth}
         \includegraphics[width=1.\textwidth]{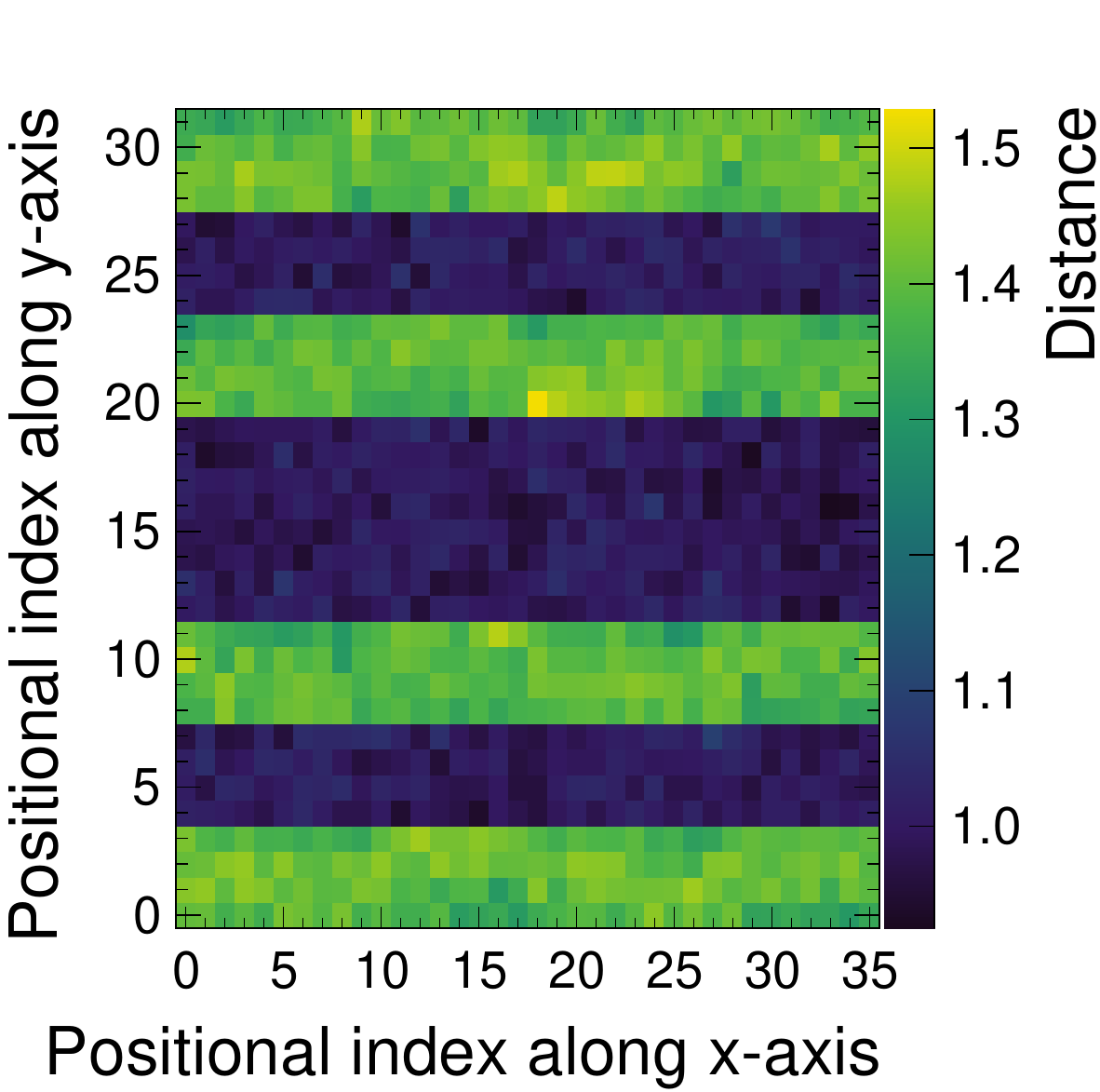}
         \caption{}
         \label{fig:5.5}
     \end{subfigure}
     \hfill
     \begin{subfigure}[b]{0.49\textwidth}
         \includegraphics[width=0.9\textwidth]{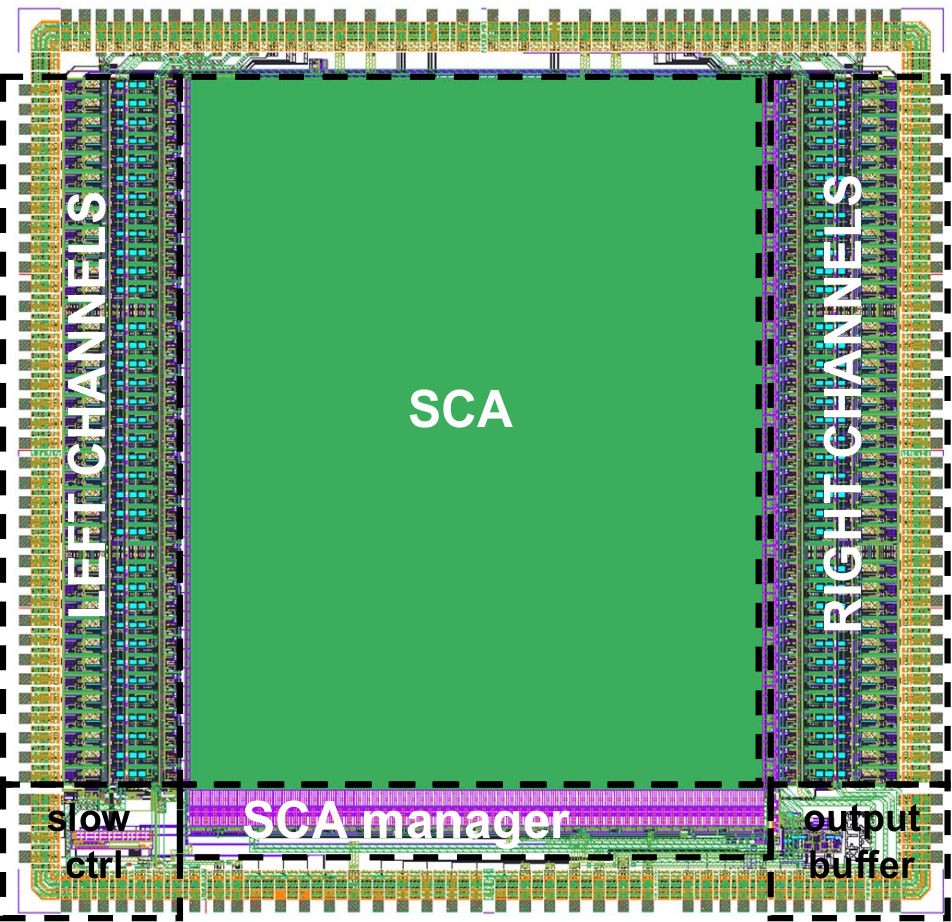}
         \caption{}
         \label{fig:5.6}
     \end{subfigure}
        \caption{
        Distance (square root of $\chi^2$) of the averaged Waveform of a pad to the average of the averaged Waveforms of all pads of the population 1 (left), and AFTER chip layout (right).}
        \label{fig:fig5.5.6}
\end{figure}

Fig.~\ref{fig:5.3} displays the average waveforms from several pads. Although the time average value of these signals is zero, they are clearly not random. The figure also reveals two distinct waveform patterns, which correspond to two separate populations of pads located on the ERAMs, as illustrated in Fig.~\ref{fig:5.5}. This partitioning is distinctly visible and divides each ASIC into two halves.

This behavior was confirmed during noise measurements from an electronics module disconnected from the detector. The 72 channels of each AFTER chip are symmetrically partitioned along an axis of the chip as illustrated in Fig.~\ref{fig:5.6}, determining whether a pad exhibits the non-random noise characteristics   of one population or the other~\cite{Baron:2008zza}.
In the following, 'pad population 1' and 'pad population 2' will designate the sets of pads interfaced with the left and right channel groups of the AFTER chip, respectively.

\begin{figure}[hbt!]
\centering
\includegraphics[scale=0.5]{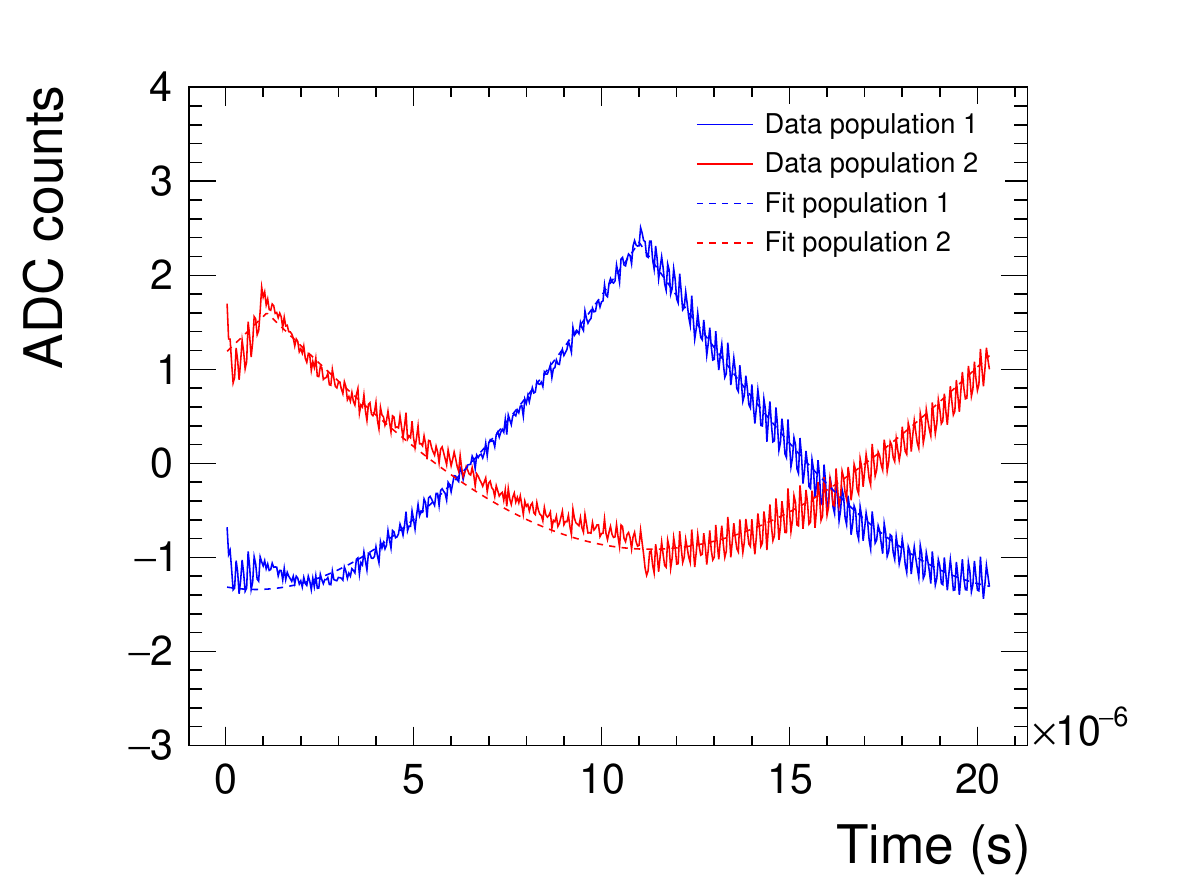}
\caption{Deterministic components of the signal ($T_p = 412$ {\rm ns},$f_e = 25 {\rm MHz})$ with the fitting function from Eq.~\eqref{equ:low_deterministic} superimposed.}
\label{5.7}
\centering
\end{figure}

Fig.~\ref{5.7} presents the average waveforms across all ERAMs, separated by pad population. To model these waveforms, we consider an absolute sine function 
\begin{equation}
f_{low}\left(t ; t_{w}, A_{\max }, t_{\max }\right)=A_{\max } \times\left(1-\frac{\pi}{2}\left|\sin \left(\pi \frac{t-t_{\max }}{t_{w}}\right)\right|\right) 
\label{equ:low_deterministic}
\end{equation}
where 
$t_w$ is the time window,
$A_{\text{max}}$ is 
the maximum amplitude
and  $t_{\max}$ is the time when the signal reaches its maximum.

Fig.~\ref{5.7} displays the deterministic components of the signal fitted using the previously defined function. The smooth dotted lines represent the fitted functions for each pad population.
The parameters of the fit are given in table~\ref{tab:parameters_low_deterministic}.

\begin{table}[htbp]
\centering
\begin{tabular}{|c|c|c|}
\hline
Pad Population & $A_{max}$ & $t_{max}$ ($\rm{\mu s}$) \\ \hline 
 1             & 2.35 & 11\\ \hline 
 2             & 1.6 & 1.1\\ \hline 
\end{tabular}
\caption{Table of parameters for the low-frequency deterministic component.}
\label{tab:parameters_low_deterministic}
\end{table}

It is clear from Fig.~\ref{5.7} that there is an other deterministic contribution at 
high frequency superimposed on the low frequency component just described.
Actually this additional component does correspond to the peak at $f_p=\frac{f_e}{4}\sim 6 \;{\rm MHz}$ displayed in Fig.~\ref{fig:Gain_02_QA_OffCenter}.
This contribution can be modeled by the function
\begin{equation}
f_{high}(t) = A_{high} \cos\left(2\pi \frac{f_e}{4}t + \pi\right) 
\label{equ:high_deterministic}
\end{equation}
with $A_{high} = 0.2 $

\subsection{White noise Component}
The system exhibits low-amplitude white noise that remains negligible at low frequencies but becomes prominent above a few {\rm MHz}, with an amplitude of approximately $28.5 \;{\rm dB}$, as shown in Fig.~\ref{fig:fig.5.2}.
As explained below, this white noise component can be modeled using a Gaussian distribution with a standard deviation of approximately $\sigma \approx 1$ ADC unit.

\section{Monte Carlo Simulations}
\label{sec:MCproductions}
With our understanding of the noise characteristics and a robust analytical description of its FFT, we can now proceed to generate noise waveforms. To construct these waveforms, we combine all relevant noise components by computing the following function for each time bin $b$ 
centered on time $t_b$:  
\begin{align}
S(t_{b}) & =  
A_{\text{low}} \left(1 - \frac{\pi}{2} \left| \sin \left( \frac{\pi (t_{b} - t_{\text{max}})}{t_w} \right) \right| \right) 
\label{Eq:6.1} \\
& + A_{\text{high}} \cos \left( 2\pi \frac{f_s}{4} t_{b} + \pi \right) 
\label{Eq:6.2}\\
& + \sum_{f} I(f) \left| H(i 2 \pi f) \right| \cos \left( \omega t_{b} + \phi_H + \phi_R \right) 
\label{Eq:6.3} \\
& + R^b_{0,\sigma} 
\label{Eq:6.4}
\end{align}
where each term corresponds to a specific contribution:
\begin{itemize}
    \item the terms in Eq.~\eqref{Eq:6.1} and Eq.~\eqref{Eq:6.2} are respectively, the low-frequency and high deterministic contributions
    (cf equations~\eqref{equ:low_deterministic} and~\eqref{equ:high_deterministic}.)
    \item the term in Eq.~\eqref{Eq:6.3}, models the AFTER chip response to a random current input
    \begin{equation*}
        I(t) = \sum_{f} I(f) \cos\left( \omega t +  \phi_R \right)
    \end{equation*}
    where $\phi_R$ is a random phase uniformly distributed between $0$ and $2\pi$.
    The term in Eq.~\eqref{Eq:6.3} is indeed the response to such a current since, by definition of
    the transfer function,  the response to a input such as
    \begin{equation*}
        I(t) = I_{0}\cos\left( \omega t \right)
    \end{equation*}
    is
    \begin{equation*}
        V(t)= I_{0} \left| H(d=i 2 \pi f) \right| \cos\left( \omega t + \phi_H \right)
    \end{equation*}
    where $| H(s=i 2 \pi f)|$ and $\Phi_H$ are the norm and phase of the transfer function (cf. Fig.~\ref{fig:fig_3.2}).

    The sum in the term in Eq.~\eqref{Eq:6.3} runs on 1000 frequencies between $\frac{f_e}{511}$ and $f_e$.

    \item in Eq.~\eqref{Eq:6.4}, $R^b_{0,\sigma}$ is a random variable drawn from a Gaussian distribution centered on 0 and of standard deviation $\sigma$ for each time bin $b$. 
    It models the noise of the output stage of the AFTER chip and external ADC. 
    An effective value of $\sigma=1.45$ was found necessary to describe the high frequency plateau.
    
\end{itemize}
The function is called to generate a collection of 511 integers, simulating a noise waveform. The parameters $\phi_R$ and $R^b_{0,\sigma}$ are the only random variables involved.

\subsection{Comparison Between Data and Simulation}
\begin{figure}[hbt!]
\centering
\includegraphics[scale=0.6]{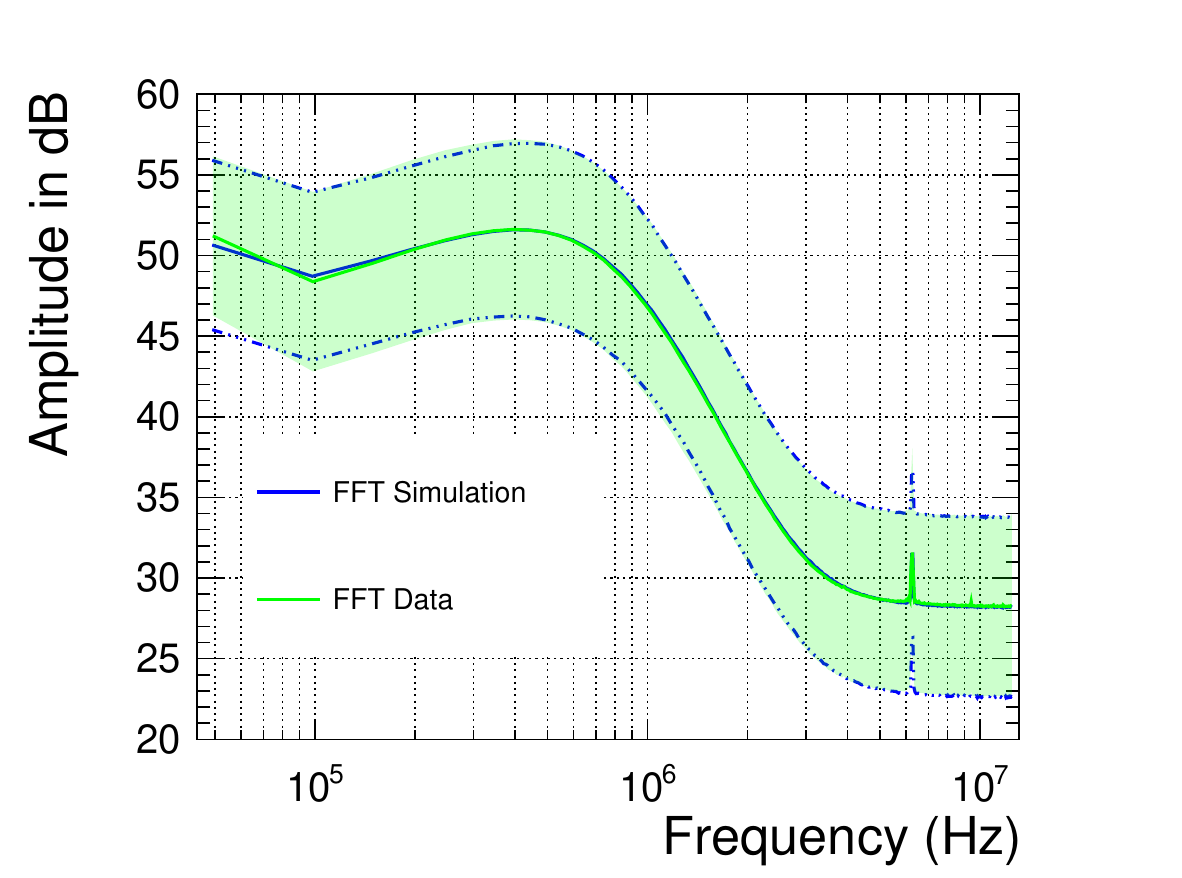}
\caption{Comparison between the means FFT spectra in simulation (blue) and data (green).}
\label{6.1}
\centering
\end{figure}

We now compare our simulation with real data. As a benchmark, we generate $10^5$ waveforms. As shown in Fig.~\ref{6.1}, we achieve excellent agreement between the mean FFT of the experimental data and the output of our simulation. Notably, the $\pm 1\sigma$ envelope also shows very good agreement
which further supports the validity of the stochastic model.

By comparing the ADC distribution, we find that the data have an RMS of 6 ADC counts, while the simulation yields an RMS of 5.8 ADC counts, indicating a very good agreement. 

\begin{figure}[hbt!]
\centering
\includegraphics[scale=0.6]{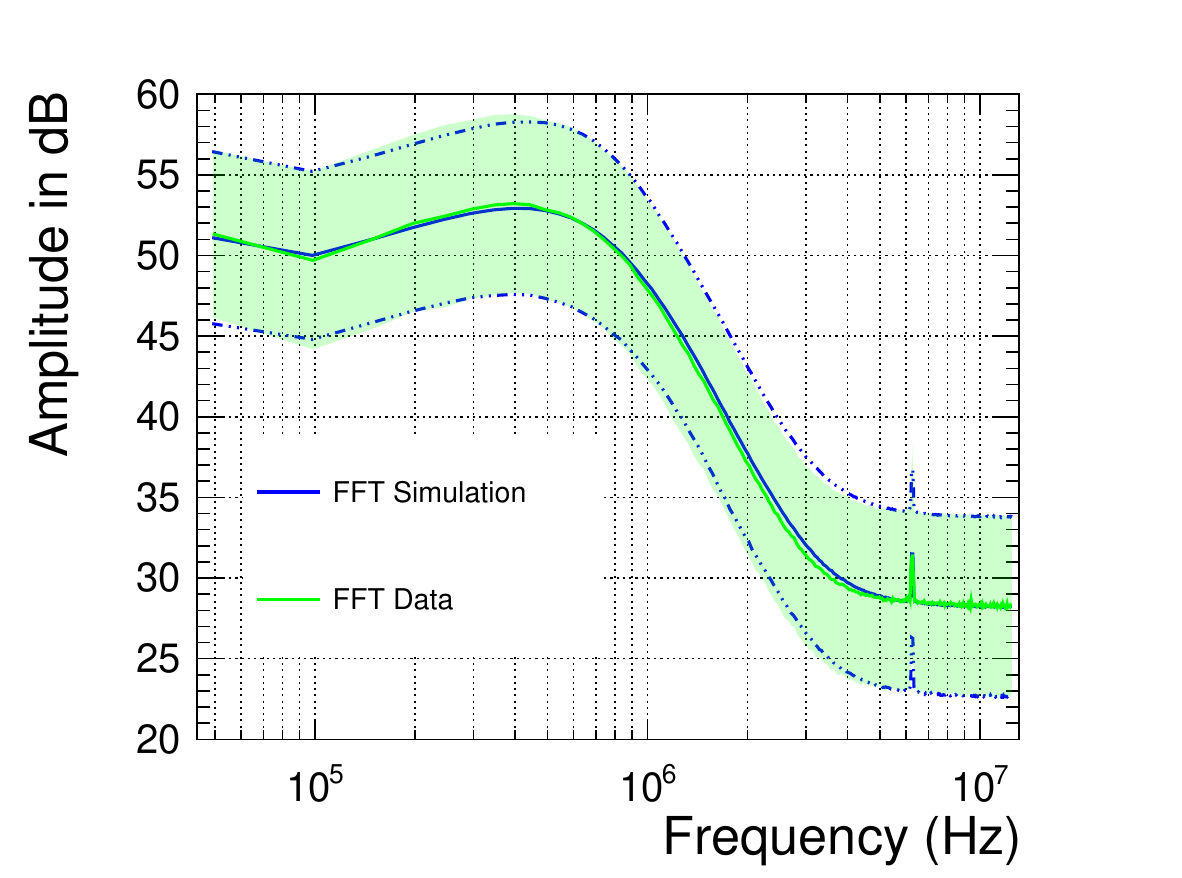}
\caption{FFT Mean for ERAM29: MC Simulations (blue) compared to real Data (green).}
\label{6.8}
\centering
\end{figure}

As shown in section~\ref{sec:FFT_analysis_section} and Fig.~\ref{fig:Gain_02_QA_OffCenter}, ERAM 29 exhibits higher amplitude noise, which can be accounted for by applying a scaling factor of $7/6$ to the term in Eq.~\eqref{Eq:6.3}. After this adjustment, we compare the mean FFT of our Monte Carlo (MC) simulations with that of the data, as illustrated in Fig.~\ref{6.8}, and find good agreement. Additionally, an analysis of the ADC distribution confirms compatibility between real data and simulation, with both having a value of 7 ADC counts.

\begin{figure}[hbt!]
     \centering
     \begin{subfigure}[b]{0.49\textwidth}
         \includegraphics[width=1.\textwidth]{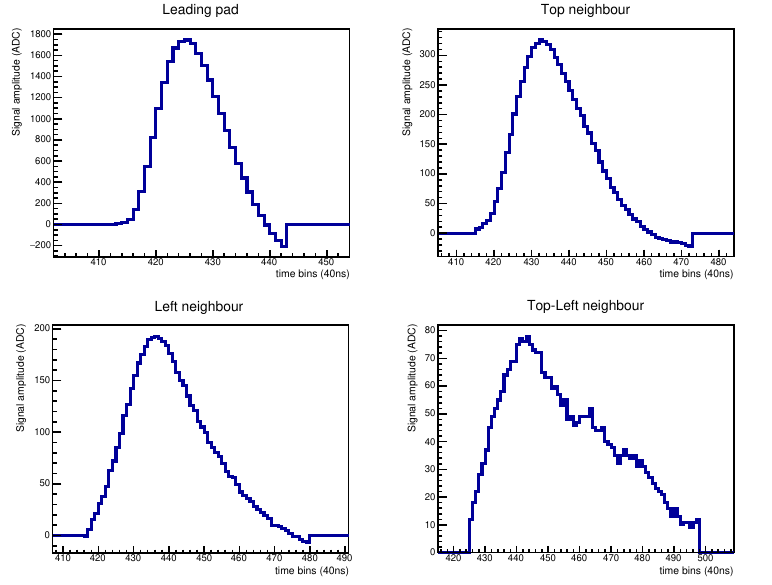}
         \caption{Waveforms in Data}
         \label{fig:WfXrayData}
     \end{subfigure}
     \hfill
     \begin{subfigure}[b]{0.49\textwidth}
         \includegraphics[width=1.\textwidth]{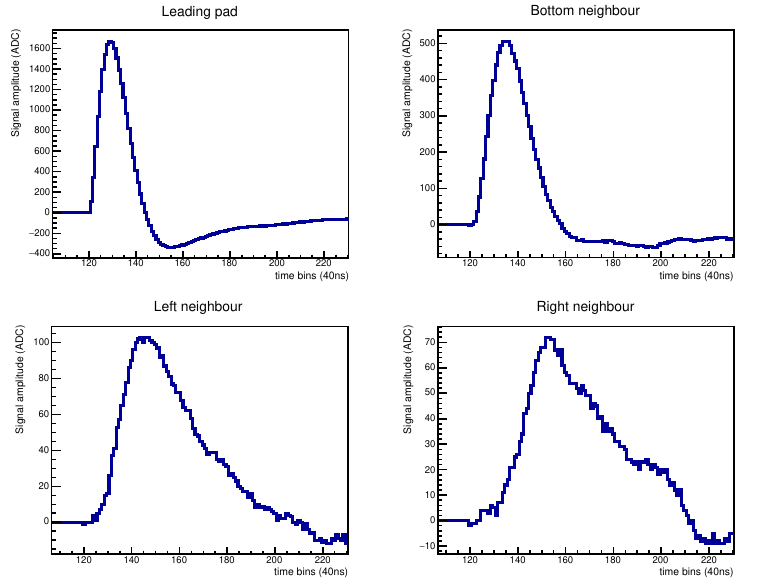}
         \caption{Waveforms in Monte Carlo Simulation}
         \label{fig:WfXraySim}
     \end{subfigure}
        \caption{Example of a comparison between X-ray data waveforms (left) and toy MC waveforms with simulated noise (right) for the leading pad and its neighboring pads. }
        \label{fig:WfXrayDataSim}
\end{figure}

Charge deposition events were generated using a toy Monte Carlo simulation, with simulated noise applied to the resulting waveforms. Fig.~\ref{fig:WfXrayDataSim} presents a comparison between waveforms from X-ray data and those from the toy MC simulation. In this example, a pad of an ERAM is scanned using an X-ray beam produced by a $^{55}$Fe source housed within a collimator. In this process, X-ray photo-electron causes an electron avalanche in the Micromegas amplification gap, directly above the targeted (leading) pad. The resulting charge is deposited onto this pad and subsequently spreads to neighboring pads. The high degree of agreement between the data and MC waveforms indicates that the real data is accurately reproduced through the implementation of the noise model. \\
In resistive Micromegas, the waveform shape plays a key role in improving reconstruction resolution, particularly in aspects such as peak position, decreasing time, and undershoot. The arrival time and magnitude of charge induced in neighboring pads depend on several factors, including the $RC$ value of the leading pad, the initial charge magnitude, and the relative position of the adjacent pads to the initial charge deposition. Therefore, accurately incorporating noise effects into waveform shapes in simulations is essential for a reliable assessment of detector performance.

\section{Conclusions}
\label{sec:conclusion}
The high-angle Time Projection Chambers of the T2K experiment are equipped with a new resistive Micromegas readout system, incorporating AFTER chips for signal processing. This study has provided a detailed analysis and characterization of the electronic noise in these detectors, leading to the development of a comprehensive noise model. The model has been validated through Monte Carlo simulations, which exhibit excellent agreement with the recorded data. All the detectors exhibit a quasi-identical and time-stable noise level. The analytical framework successfully describes the observed noise characteristics, offering a robust tool for studying systematic effects in signal processing. These results will be incorporated into the HA-TPC simulation, contributing to a more precise understanding and improved estimation of the detector resolution.
\section*{Acknowledgements}
The authors would like to acknowledge the work performed by the student interns,  S. Ramatchandirin , R. Fourquet, M. Dounas, funded by IRFU, CEA, Université Paris-Saclay, Gif-sur-Yvette, France. We acknowledge the support of CEA and CNRS/IN2P3, France; DFG, Germany; INFN, Italy.\\
This work was supported by P2IO LabEx (ANR-10-LABX-0038 – Project “BSMNu”) in the framework "Investissements d’Avenir" (ANR-11-IDEX-0003-01), managed by
the Agence Nationale de la Recherche (ANR), France. In addition, the participation of individual researchers and institutions has
 been further supported by H2020 Grant No. RISE-GA822070-JENNIFER2 2020.\\
The research leading to these results has received funding from the Spanish Ministry of Science and Innovation \textnormal{PID2022-136297NB\_I00/AEI\allowbreak/10.13039\allowbreak/501100011033\allowbreak/FEDER,\allowbreak\ UE}. IFAE is partially funded by the CERCA program of the Generalitat de Catalunya. With the support from the Secretariat for Universities and Research of the Ministry of Business and Knowledge of the Government of Catalonia and the European Social Fund (2022FI$_{B}$ 00336).




\bibliographystyle{elsarticle-num}
\bibliography{bibliography}

\begin{thebibliography}{1}
\expandafter\ifx\csname url\endcsname\relax
  \def\url#1{\texttt{#1}}\fi
\expandafter\ifx\csname urlprefix\endcsname\relax\def\urlprefix{URL }\fi
\expandafter\ifx\csname href\endcsname\relax
  \def\href#1#2{#2} \def\path#1{#1}\fi

\bibitem{T2K:2011qtm}
K.~Abe, et~al., {The T2K Experiment}, Nucl. Instrum. Meth. A 659 (2011) 106--135.
\newblock \href {http://arxiv.org/abs/1106.1238} {\path{arXiv:1106.1238}}, \href {https://doi.org/10.1016/j.nima.2011.06.067} {\path{doi:10.1016/j.nima.2011.06.067}}.

\bibitem{Attie:2022smn}
D.~Atti\'e, et~al., {Analysis of test beam data taken with a prototype of TPC with resistive Micromegas for the T2K Near Detector upgrade}, Nucl. Instrum. Meth. A 1052 (2023) 168248.
\newblock \href {http://arxiv.org/abs/2212.06541} {\path{arXiv:2212.06541}}, \href {https://doi.org/10.1016/j.nima.2023.168248} {\path{doi:10.1016/j.nima.2023.168248}}.

\bibitem{Ambrosi:2023smx}
L.~Ambrosi, et~al., {Characterization of charge spreading and gain of encapsulated resistive Micromegas detectors for the upgrade of the T2K Near Detector Time Projection Chambers}, Nucl. Instrum. Meth. A 1056 (2023) 168534.
\newblock \href {http://arxiv.org/abs/2303.04481} {\path{arXiv:2303.04481}}, \href {https://doi.org/10.1016/j.nima.2023.168534} {\path{doi:10.1016/j.nima.2023.168534}}.

\bibitem{Baron:2008zza}
P.~Baron, D.~Calvet, E.~Delagnes, X.~de~la Broise, A.~Delbart, F.~Druillole, E.~Monmarthe, E.~Mazzucato, F.~Pierre, M.~Zito, {AFTER, an ASIC for the readout of the large T2K time projection chambers}, IEEE Trans. Nucl. Sci. 55 (2008) 1744--1752.
\newblock \href {https://doi.org/10.1109/TNS.2008.924067} {\path{doi:10.1109/TNS.2008.924067}}.

\bibitem{Attie:2021yeh}
D.~Atti\'e, et~al., {Characterization of resistive Micromegas detectors for the upgrade of the T2K Near Detector Time Projection Chambers}, Nucl. Instrum. Meth. A 1025 (2022) 166109.
\newblock \href {http://arxiv.org/abs/2106.12634} {\path{arXiv:2106.12634}}, \href {https://doi.org/10.1016/j.nima.2021.166109} {\path{doi:10.1016/j.nima.2021.166109}}.

\bibitem{Attie:2019hua}
D.~Atti\'e, et~al., {Performances of a resistive Micromegas module for the Time Projection Chambers of the T2K Near Detector upgrade}, Nucl. Instrum. Meth. A 957 (2020) 163286.
\newblock \href {http://arxiv.org/abs/1907.07060} {\path{arXiv:1907.07060}}, \href {https://doi.org/10.1016/j.nima.2019.163286} {\path{doi:10.1016/j.nima.2019.163286}}.

\bibitem{Calvet:2018lac}
D.~Calvet, {Back-End Electronics Based on an Asymmetric Network for Low Background and Medium- Scale Physics Experiments}, IEEE Trans. Nucl. Sci. 66~(7) (2018) 998--1006.
\newblock \href {http://arxiv.org/abs/1806.07618} {\path{arXiv:1806.07618}}, \href {https://doi.org/10.1109/TNS.2018.2884617} {\path{doi:10.1109/TNS.2018.2884617}}.

\end{thebibliography}

\end{document}